\documentclass[lettersize,journal]{IEEEtran}
\usepackage{amsmath,amsfonts}
\usepackage{algorithmic}
\usepackage{algorithm}
\usepackage{array}
\usepackage[caption=false]{subfig}
\usepackage{textcomp}
\usepackage{stfloats}
\usepackage{url}
\usepackage{verbatim}
\usepackage{graphicx}
\usepackage[nolist]{acronym}
\usepackage{cite}
\usepackage{svg}
\usepackage{soul}
\usepackage{color}
\usepackage{xcolor}
\usepackage{tikz}
\usepackage{pgfplots}
\usepackage{comment}
\pgfplotsset{compat=1.18}
\definecolor{mycolor1}{rgb}{0.50000,0.00000,0.80000}%
\definecolor{mycolor2}{rgb}{0.47000,0.67000,0.19000}%
\definecolor{cpm100}{rgb}{0.00,0.00,0.00}%
\definecolor{cpm200}{rgb}{0.50,0.50,0.50}%
\definecolor{cpm400}{rgb}{0.70,0.70,0.70}%
\definecolor{random}{rgb}{0.55,0.30,0.65}%
\definecolor{fixedtwo}{rgb}{0.45,0.65,0.25}%
\usepackage[normalem]{ulem}
\usepackage{soulutf8}

\usetikzlibrary{fit,backgrounds}
\usepgfplotslibrary{groupplots,statistics}
\usepackage{standalone}
\usetikzlibrary{positioning,calc,arrows.meta,fit}

\pgfplotsset{
    myplotstyle/.style={
        width=9.5cm, 
        height=7cm,
        grid=both,
        major grid style={gray!30, solid},
        minor grid style={gray!10, dashed},
        xlabel={$D$ [m]},
        xmin=10, xmax=500,
        tick label style={font=\huge},
        label style={font=\huge},
        title style={at={(0.5, -0.28)}, anchor=north, font=\huge}
    }
}

\newcommand{\rev}[1]{\textcolor{black}{#1}}
\newcommand{\revbis}[1]{\textcolor{black}{#1}}

\newcommand{\revtris}[1]{\textcolor{black}{#1}}

\newcommand{\revremove}[1]{}

\begin{acronym} 
\acro{3GPP}{Third Generation Partnership Project}
\acro{2D}{two-dimensional}
\acro{2G}{second generation}
\acro{3G}{third generation}
\acro{4G}{fourth generation}
\acro{5G}{fifth generation}
\acro{6G}{sixth generation}
\acro{5GAA}{5G Automotive Association}
\acro{5GS}{5G system}
\acro{ACC}{adaptive cruise control}
\acro{AD}{autonomous driving}
\acro{ADAS}{advanced driver assistance systems}
\acro{AF}{application function}
\acro{AGC}{automatic gain control}
\acro{AGV}{automated guided vehicle}
\acro{AI}{artificial intelligence}
\acro{AIFS}{arbitration inter-frame space}
\acro{AMC}{adaptive modulation and coding}
\acro{ANRV}{average normalized received value of information} 
\acro{AoI}{age of information}
\acro{API}{application programming interface}
\acro{AS}{application server}
\acro{AD}{autonomous driving}
\acro{AUTOSAR}{Automotive Open System}
\acro{AWGN}{additive white Gaussian noise}
\acro{BEP}{beacon error probability}
\acro{BM-SC}{broadcast multicast service center}
\acro{BP}{beacon periodicity}
\acro{B-CSA}{broadcast \ac{CSA}}
\acro{BSM}{basic safety message}
\acro{BSS}{basic service set}
\acro{BS}{base station}
\acro{BTP}{Basic Transport Protocol}
\acro{BWP}{bandwidth part}
\acro{C-ACC}{cooperative adaptive cruise control}
\acro{C-GLOSA}{cooperative GLOSA}
\acro{C-ITS}{cooperative-intelligent transport system}
\acro{C-NOMA}{cooperative \ac{NOMA}}
\acro{C-V2X}{cellular-\ac{V2X}}
\acro{C2C-CC}{Car 2 Car Communication Consortium}
\acro{CA}{collision avoidance}
\acro{CACC}{cooperative adaptive cruise control}
\acro{CAM}{cooperative awareness message}
\acro{CAV}{connected and autonomous vehicle}
\acro{CBR}{channel busy ratio}
\acro{CC}{cruise control}
\acro{CCA}{clear channel assessment}
\acro{CCAM}{cooperative, connected and automated mobility}
\acro{CCDF}{complementary cumulative distribution function}
\acro{CDF}{cumulative distribution function}
\acro{CDMA}{code-division multiple access}
\acro{CEPT}{European Conference of Postal and Telecommunications Administrations}
\acro{CoMP}{coordinated multi-point}
\acro{CMM}{co-channel coexistence mitigation methods}
\acro{CN}{core network}
\acro{CP-OFDM}{cyclic prefix orthogonal frequency-division multiplexing}
\acro{CP}{collective perception}
\acro{CPM}{collective perception message}
\acro{CPS}{collective perception service}
\acro{CR}{channel occupancy ratio}
\acro{CRDSA}{contention resolution diversity slotted ALOHA}
\acro{CRLB}{Cram\'er-Rao Lower Bound }
\acro{CSA}{coded-slotted ALOHA}
\acro{CSD}{cyclic shift diversity}
\acro{CSI}{channel state information}
\acro{CSIT}{channel state information at the transmitter}
\acro{CSIR}{channel state information at the receiver}
\acro{CSMA/CA}{carrier sense multiple access with collision avoidance}
\acro{CSMS}{cyber security management system}
\acro{CSSR}{candidate single-subframe resource}
\acro{CTS-To-Self}{Clear-To-Send-To-Self}
\acro{DA}{data age}
\acro{D2D}{device-to-device}
\acro{DC}{duty cycle}
\acro{DCC}{decentralized congestion control}
\acro{DCF}{distributed coordination function}
\acro{DCI}{downlink control
information}
\acro{DCM}{dual carrier modulation}
\acro{DENM}{decentralized environmental notification message}
\acro{DMRS}{demodulation reference signal}
\acro{DoS}{denial-of-service}
\acro{DDoS}{distributed denial-of-service}
\acro{DOT}{Department of Transportation}
\acro{DRX}{discontinuous reception}
\acro{DSRC}{Dedicated Short Range Communication}
\acro{DS}{dynamic scheduling}
\acro{DT}{digital twin}
\acro{DtS}{direct-to-satellite}
\acro{DtS-IoT}{direct-to-satellite Internet of Things}
\acro{EC}{European Commission}
\acro{ECP}{extended cyclic prefix}
\acro{EDCA}{enhanced distributed coordination access}
\acro{EDGE}{Enhanced Data rates for GSM Evolution}
\acro{EED}{eEnd-to-end delay}
\acro{eMBMS}{evolved Multicast Broadcast Multimedia Service}
\acro{eMBB}{enhanced mobile broadband}
\acro{eNodeB}{evolved NodeB}
\acro{EPC}{Evolved Packet Core}
\acro{EPS}{Evolved Packet System}
\acro{ES}{energy signal}
\acro{ETSI}{European Telecommunications Standards Institute}
\acro{eV2X}{enhanced V2X}
\acro{FANET}{flying ad hoc network}
\acro{FCC}{Federal Communications Commission}
\acro{FCD}{floating car data}
\acro{FD}{full-duplex}
\acro{FDD}{frequency division duplex}
\acro{FDMA}{frequency division multiple access}
\acro{FEC}{forward error correction}
\acro{FR1}{frequency range 1}
\acro{FR2}{frequency range 2}
\acro{GapRe}{gap sensing and resource re-selection}
\acro{GLOSA}{green light optimal speed advisory}
\acro{GNSS}{global navigation satellite system}
\acro{GNW}{GeoNetworking}
\acro{GPRS}{general packet radio service }
\acro{GPS}{global positioning system}
\acro{HARQ}{hybrid automatic repeat request}
\acro{HD}{high-definition}
\acro{HPC}{high performance computing}
\acro{HRLLC}{hyperreliable
and low-latency communication}
\acro{HSDPA}{High Speed Downlink Packet Access}
\acro{HSPA}{High Speed Packet Access}
\acro{IBE}{in-band emission}
\acro{ICI}{inter-carrier interference}
\acro{IDMA}{interleave-division multiple access}
\acro{IEEE}{Institute of Electrical and Electronics Engineers}
\acro{IM}{index modulation}
\acro{IoT}{Internet of Things}
\acro{IP}{Internet Protocol}
\acro{IPG}{inter-packet gap}
\acro{IR}{incremental redundancy}
\acro{IRS}{intelligent reflective surface}
\acro{ISAC}{integrated sensing and communication}
\acro{ISI}{inter-symbol interference}
\acro{ISM}{industrial, scientific, and medical}
\acro{ISO}{international standardization organization}
\acro{ITU-R}{International Telecommunication Union Radiocommunication Sector}
\acro{ITS}{intelligent transport system}
\acro{i.i.d.}{independent identically distributed}
\acro{KPI}{key performance indicator}
\acro{LBT}{listen-before-talk}
\acro{LDPC}{low-density parity-check}
\acro{LDS}{low-density spreading}
\acro{LEO}{low Earth orbit}
\acro{LIDAR}{light detection and ranging}
\acro{LLR}{log-likelihood ratio}
\acro{LOD}{Linked Open Data}
\acro{LOS}{line-of-sight}
\acro{LoRa}{Long Range}
\acro{LoRa-CSS}{long range chirp spread spectrum}
\acro{LoRaWAN}{Long Range Wide Area Network}
\acro{LPWAN}{low-power wide-area network}
\acro{LR-FHSS}{long range frequency hopping spread spectrum}
\acro{LTE}{long term evolution}  
\acro{LTE-D2D}{\ac{LTE} with \ac{D2D} communications}  
\acro{LTE-LAA}{LTE license assisted access}
\acro{LTE-V2X}{long-term-evolution-vehicle-to-anything} 
\acro{LTE-V2V}{\ac{LTE}-\ac{V2V}}
\acro{LwM2M}{lightweight machine-to-machine} 
\acro{MAC}{medium access control}
\acro{MANET}{mobile ad hoc network}
\acro{MBMS}{Multicast Broadcast Multimedia Service}
\acro{MBMS-GW}{MBMS Gateway}
\acro{MBSFN}{MBMS Single Frequency Network}
\acro{MCE}{Multi-cell Coordination Entity}
\acro{MEC}{mobile edge computing}
\acro{MCM}{maneuver coordination message}
\acro{MCS}{modulation and coding scheme}
\acro{MIMO}{multiple input multiple output}
\acro{ML}{machine learning}
\acro{mMTC}{massive machine-type communications}
\acro{MRC}{maximum ratio combining}
\acro{MRD}{maximum reuse distance}
\acro{MUD}{multi-user detection}
\acro{NAV}{network allocation vector}
\acro{NB-IoT}{Narrowband IoT}
\acro{NEF}{network exposure function}
\acro{NGSI-LD}{next generation service interfaces - linked data}
\acro{NGV}{Next Generation V2X}
\acro{NHTSA}{National Highway Traffic Safety Administration}
\acro{NIS2}{network and information systems directive}
\acro{NLOS}{non-line-of-sight}
\acro{NOMA}{non-orthogonal multiple access}
\acro{NOMA-MCD}{\ac{NOMA}-mixed centralized/distributed}
\acro{NR}{New Radio}
\acro{NR-U}{NR unlicensed}
\acro{NTN}{non-terrestrial network}
\acro{OBU}{onboard unit}
\acro{OCB}{outside of the context of a basic service set}
\acro{OEM}{original equipment manufacturer}
\acro{OFDM}{orthogonal frequency-division multiplexing}
\acro{OFDMA}{orthogonal frequency-division multiple access}
\acro{OMA}{orthogonal multiple access}
\acro{OTFS}{orthogonal time frequency space}
\acro{pdf}{probability density function}
\acro{PAPR}{peak to average power ratio}
\acro{PCF}{policy control function}
\acro{PCM}{platoon control message}
\acro{PD}{packet delay}
\acro{PEP}{pairwise error probability} 
\acro{PER}{packet error rate}
\acro{PIAT}{packet inter-arrival time}
\acro{PIR}{packet inter reception}
\acro{PLS}{physical layer security}
\acro{PPDU}{physical layer protocol data unit}
\acro{PRB}{physical resource block}
\acro{PSCCH}{Physical Sidelink Control Channel}
\acro{PSDU}{physical layer convergence protocolservice data unit}
\acro{PSFCH}{Physical Sidelink Feedback channel}
\acro{PSSCH}{Physical Sidelink Shared channel}
\acro{PHY}{physical}
\acro{PL}{path loss}
\acro{PLMN}{Public Land Mobile Network}
\acro{PPPP}{ProSe Per-Packet Priority}
\acro{PPPR}{ProSe Per-Packet Reliability}
\acro{ProSe}{Proximity-based Services}
\acro{PRR}{packet reception ratio}
\acro{QAM}{quadrature amplitude modulation}
\acro{QC-LDPC}{quasi-cyclic \ac{LDPC}}
\acro{QoS}{quality of service}
\acro{QPSK}{quadrature phase shift keying}
\acro{RAN}{radio access network}
\acro{RAT}{radio access technology}
\acro{RB}{resource block}
\acro{RCS}{radar cross-section}
\acro{RE}{resource element}
\acro{RF}{radio frequency}
\acro{RIS}{reconfigurable intelligent  surfaces}
\acro{RL}{reinforcement learning}
\acro{RNN}{recurrent neural network}
\acro{RR}{radio resource}
\acro{RRC}{radio resource control}
\acro{RRI}{resource reservation interval}
\acro{RSRP}{reference signal received power}
\acro{RSSI}{Received Signal Strength Indicator}
\acro{RSU}{road side unit} 
\acro{SAE}{Society of Automotive Engineers}
\acro{SAI}{Service Area Identifier}
\acro{SB-DS}{sensing-based dynamic scheduling}
\acro{SB-SPS}{sensing-based semi-persistent scheduling}
\acro{SC-FDMA}{single carrier frequency division multiple access}
\acro{SC-PTM}{Single Cell Point To Multipoint}
\acro{SCS}{subcarrier spacing}
\acro{SCMA}{sparse-code multiple access}
\acro{SI}{self-interference}
\acro{SIC}{successive interference cancellation}
\acro{SIFS}{short inter-frame space}
\acro{S-UE}{scheduling UE}
\acro{SCI}{sidelink control information}
\acro{SDG}{sustainable development goals}
\acro{SDN}{software-defined networking}
\acro{SF}{spreading factor}
\acro{SFFT}{symplectic finite Fourier transform}
\acro{SHINE}{simulation platform for heterogeneous interworking networks}
\acro{SINR}{signal-to-interference-plus-noise ratio}
\acro{SL}{sidelink}
\acro{SNR}{signal-to-noise ratio}
\acro{SRS}{sounding reference signal}
\acro{SPS}{semi-persistent scheduling}
\acro{STBC}{space time block codes}
\acro{SV}{smart vehicle}
\acro{TB}{transport block}
\acro{TBC}{time before change}
\acro{TBE}{time before evaluation}
\acro{TDD}{time-division duplex}
\acro{TDMA}{time-division multiple access}
\acro{TR}{transmission range}
\acro{TS}{Technical Specification}
\acro{TTI}{transmission time interval}
\acro{UAV}{uncrewed aerial vehicle}
\acro{UE}{User Equipment}
\acro{UD}{update delay}
\acro{UDD}{urban digital development}
\acro{UMTS}{universal mobile telecommunications system}
\acro{URI}{Uniform Resource Identifier} 
\acro{URLLC}{ultra-reliable and ultra-low latency communications}
\acro{USIM}{Universal Subscriber Identity Module}
\acro{UTDOA}{uplink time difference of arrival}
\acro{UWAC}{underwater acoustic communications}
\acro{V2C}{vehicle-to-cellular} 
\acro{V2N}{vehicle-to-network}
\acro{V2I}{vehicle-to-infrastructure} 
\acro{V2P}{vehicle-to-pedestrian}
\acro{V2R}{vehicle-to-roadside}
\acro{V2V}{vehicle-to-vehicle} 
\acro{V2X}{vehicle-to-everything} 
\acro{VoI}{value of information}
\acro{VANET}{vehicular ad hoc network}
\acro{VAM}{VRU awareness message}
\acro{VRU}{vulnerable road user}
\acro{WAVE}{wireless access in vehicular environment}
\acro{WBSP}{wireless blind spot probability}
\acro{XML}{extensible markup language}
\acro{SL}{sidelink}
\acro{ECC}{Electronic Communications Committee}
\end{acronym}

\begin{document}

\title{The Role of Collective Perception and \\5G NR-V2X Sidelink in Road Safety}

\author{Vittorio Todisco, Mattia Andreani, Maria Luisa Merani and Alessandro Bazzi
\thanks{Vittorio Todisco and Alessandro Bazzi are with the \emph{DEI Department, University of Bologna, Bologna, Italy,} and the \emph{National Laboratory of Wireless Communications (WiLab), CNIT}.}
\thanks{Mattia Andreani and Maria Luisa Merani are with the \emph{Department of Engineering ``Enzo Ferrari'', University of Modena and Reggio Emilia, Modena, Italy,} and \emph{CNIT}.}
}


\maketitle

\begin{abstract}
\revbis{Vehicles and roadside infrastructure are increasingly equipped with sensors capable of perceiving their surroundings. Sharing this information through vehicle-to-everything (V2X) communications is a key enabler of Day-2 applications and is supported by the ETSI collective perception service (CPS). While CPS is expected to play a fundamental role in future intelligent transportation systems, its operation may significantly increase channel load, posing challenges in terms of radio resource utilization, communication reliability, and information management. This paper reviews the current status of CPS standardization and investigates its impact in dense deployment scenarios where connected vehicles communicate through fifth-generation (5G) New Radio-V2X (NR-V2X) sidelink (SL) communications. The main contribution is a realistic evaluation of communication reliability, latency, channel occupancy, and information usefulness under different object-selection strategies for collective perception messages. The analysis is conducted through a network-level simulation framework integrating empirical object traces derived from real-world datasets, thereby avoiding the limitations of synthetic traffic models. Results show that perception message generation and radio access mechanisms are tightly coupled and should be jointly designed to maximize the benefits of collective perception services.}
\end{abstract} 


\section{Introduction}

\IEEEPARstart{I}n recent years, road infrastructure and vehicles have undergone a radical transformation, driven by the advent of electric propulsion systems and autonomous vehicles. In parallel, an increased attention to sustainable mobility has emerged,
spurring the usage of individual means of transportation alternative to passenger cars, such as bicycles and e-scooters. Still, road fatalities remain one of the worst scourges of our time. Increasing road safety requires a holistic approach that spans from urban redesign for pedestrians and cyclists to the adoption of all technologies
that contribute to sheltering the travelers. In this arena, the European Union is among the frontrunners, formally stating the long-term goal to move as close as possible to zero fatalities in road transport by $2050$.

On the technological front,
autonomous or highly automated vehicles and artificial intelligence tools can greatly reduce human errors, contributing to proactive crash prevention in many ways. Yet, a further technology brick is mandatory: vehicles should
exhibit cooperation capabilities to overcome the isolated, egocentric view that characterizes even the most sophisticated intelligent vehicle. 
In turn, cooperation relies on connectivity, 
enabled by \ac{V2V}, \ac{V2I}, \ac{V2N}, \ac{V2P}, and, more generally, \ac{V2X} communications.

At present, the deployment of what is known in Europe as the \ac{C-ITS}\acused{ITS} is still in its infancy. The scenario may change quickly with the advent of collective perception, where vehicles \rev{equipped with \acp{OBU}, as well as} \acp{RSU}, broadcast not only their own status but also what they perceive in the surrounding environment.
\revbis{In the literature, existing works on \ac{CPS} primarily investigate the impact of perception traffic
on radio resource usage by relying on synthetic traces of detected  objects and earlier versions of
the \ac{CPM} generation rules  (e.g., ETSI TS 103.324 V2.1.1 or TR 103.562 V2.1.1) \cite{bib:WolfCPS2015,bib:WolfCPS2016,bib:Elche_Redundancy_Mitigation,bib:Festag_Redundancy_Mitigation}. 
As a
result, two key aspects remain insufficiently understood. First, there is limited knowledge of how
many objects are perceived by real vehicles and how this affects the \ac{CPM} traffic characteristics.
Second, perception and communication are treated separately, whereas the CPS effectiveness
depends on their interaction, especially in light of the most recent standardization efforts. ETSI
specifications (such as those in TS 103.324) are moving toward adaptive message construction mechanisms based on the \ac{VoI}, not reflected in prior works.
The present study addresses these gaps by integrating object traces from a real autonomous
vehicle dataset into a simulation framework, and examining multiple strategies to build perception messages via \ac{VoI}-based or \ac{VoI}-agnostic strategies. 
The novel contributions of this work can be summarized as follows:
\begin{itemize}
    \item An updated and contextualized view of \ac{CPS} standardization, highlighting recent developments and open challenges; 
    \item A realistic characterization of \ac{CPS} traffic by leveraging real-world perception data; 
    \item An assessment of the 5G NR-V2X \ac{SL} performance under different strategies for selecting the objects to embed in the perception messages, including \ac{VoI}-inspired approaches aligned with recent ETSI developments. The evaluation is carried out in terms of reliability, timeliness, channel congestion, and value of the received information. 
\end{itemize}
The remainder of this article first explains the vehicular communication alternatives and the foreseen evolution
of safety services, then discusses collective perception, and finally provides some
key performance results when 5G NR-V2X SL is the enabling radio access technology, identifying some open challenges in this research area.
%
%
%
} 

\section{The Roadmap to V2X-based Safety}\label{sec:roadmap}

\subsection{Organizations and Standardization Bodies}

The importance of road safety is reflected by the number of organizations exerting an influence to foster the adoption of vehicular connectivity and build a secure traveling environment in different world regions. 

Wireless access technologies form the foundation of \ac{C-ITS}, and the bodies that release standards in this field are the \ac{IEEE} and the \ac{3GPP}. IEEE developed the $802.11$p standard for vehicular communications during the first decade of this century, and its enhancement $802.11$bd in the latest years; both stemmed from the original $802.11$ specifications for wireless local area networks.
Since $2016$, \ac{3GPP} introduced the alternative group of solutions called \ac{C-V2X}, first providing support to vehicular communication via LTE-V2X, and later adding  
the \ac{5G} \ac{NR}-V2X as a further option. 

As regards the standards that define the functional aspects, in Europe, the main reference organization is the \ac{ETSI}, whose \ac{ITS} technical committee develops documents covering the whole communication stack on which the ITS applications rely. In the US, a similar role is played by the \ac{SAE}, which has the advancement in connected and automated mobility systems among its core functions. Associations also play an important role. Among them, the CAR2CAR Communication Consortium (C2C-CC) and the C-ROADS Platform are key European players supporting the Vision Zero objective through the deployment of \ac{C-ITS}. C2C-CC focuses on V2X communication standards, interoperability specifications, and deployment strategies among automotive stakeholders; C-ROADS primarily includes road authorities and operators involved in harmonizing and testing C-ITS services across borders. A further major actor is the cross-regional industry consortium \ac{5GAA}, which has strongly promoted C-V2X-based solutions through white papers and field tests over the past decade.

\subsection{Enabling Communication Technologies}
From the standpoint of access technologies, the first introduced solution is \ac{IEEE} $802.11$p, ratified in $2010$ as a variant of the specifications for wireless local area networks, adapted for the highly dynamic vehicular environment; in Europe, it is the basis for ITS-G5. It operates in the $5.9$~GHz \ac{ITS} band with $10$~MHz channels, supporting raw data rates ranging from~$3$ to $27$~Mbps. The channel access mechanism is based on a listen-before-talk principle, with unicast and broadcast capabilities. In $2023$, IEEE $802.11$bd was defined to enhance data throughput, reliability, and latency; it supports wider channel bandwidth (up to $40$~MHz), higher-order modulation schemes, and new coding options, with a peak data rate of $200$~Mbps \cite{bib:evolutionV2Xstandards}. Crucially, $802.11$bd, now part of $802.11$-$2024$ and ITS-G5, ensures interoperability with $802.11$p, enabling a smooth transition and coexistence in mixed vehicular networks. To date, more than $2$ million ITS-G5-equipped vehicles have been sold in Europe.

In recent years, \ac{3GPP} introduced LTE-V2X in Release~$14$ (\ac{3GPP} TS 36.321) and enhanced it in Release~$15$. It natively supports two communication types: direct \ac{SL} communications over the PC5 interface (for \ac{V2V}, \ac{V2I}, and \ac{V2P}) and network-based communications over the Uu interface (for \ac{V2N}). To address the stringent requirements of connected and automated driving, \ac{3GPP} later specified \ac{5G} \ac{NR}-\ac{V2X} in Release $16$ (\ac{3GPP} TS 38.321). \ac{5G} \ac{NR}-\ac{V2X} redesigns 
LTE-V2X 
to 
support 
periodic and aperiodic traffic, as well as a broader set of communication types (unicast, groupcast, and broadcast). \ac{5G} \ac{NR}-\ac{V2X} \ac{SL} employs a CP-OFDM waveform, organizing the time domain in synchronous time slots.
In the frequency domain, the available bandwidth is divided into subchannels, consisting of contiguous \acp{RB} which constitute the smallest allocation units. Each transmission is further characterized by a specific \ac{MCS}, which determines the combination of modulation order and coding rate used to map bits to physical resources. \ac{5G}~\ac{NR}-\ac{V2X} \ac{SL} supports two resource allocation modes: in Mode $1$, resource scheduling is managed by the cellular base station (gNB), while in Mode $2$, vehicles autonomously select resources from a pre-configured pool, enabling the operation outside of cellular coverage. In Mode $2$, users select resources after performing a channel sensing phase, then access them according to two alternative mechanisms for distributed resource allocation: \ac{SB-SPS} and \ac{SB-DS}. With \ac{SB-SPS}, a vehicle reserves radio resources at regular intervals for a predefined number of consecutive transmissions. After completing these transmissions, it either reselects new resources with probability 1-$P$ or retains the previous allocation with probability $P$. If a newly generated message does not fit within the reserved resources, a new selection is triggered. \ac{SB-SPS} is particularly suitable for constant-size, periodic traffic. In contrast, \ac{SB-DS} allows vehicles to dynamically select resources based on current traffic conditions, making it more appropriate for irregular or event-driven messages, such as emergency alerts. However, since transmissions in \ac{SB-DS} are not generally protected by prior reservations, their reliability is typically lower than that of \ac{SB-SPS}. In China and the US, \ac{C-V2X} technology is currently being selected as the only solution, with some uncertainty about whether LTE or 5G NR will be adopted.

\subsection{The Progression}

Safety services, built on top of either \ac{IEEE} or \ac{3GPP} radio technologies, are evolving through three main phases, commonly referred to as Day~$1$, Day~$2$, and Day~$3$ of cooperative driving. In \emph{Day~1}, the paradigm of the connected road user is \rev{\emph{``I share who and where I am, and how my dynamic attributes are changing''.}} The broadcasting of messages to stations within radio range \rev{augments the vehicles' perception of} the environment and builds cooperative awareness. Examples of Day~$1$ messages include the \acp{CAM}, standardized by \ac{ETSI}, which convey vehicle mobility updates with a generation rate that adapts to changes in position, speed, and heading, and the \acp{DENM}, which carry event-driven information such as the presence of roadworks or a recently occurred accident. A major challenge for Day~1 safety services is that their efficacy depends strongly on the penetration rate of connected vehicles and infrastructures. Moreover, although specific messages broadcast by \acp{VRU}, called \acp{VAM}, have been defined, most pedestrians and cyclists are expected to remain unconnected, revealing the intrinsic limitation of cooperative awareness.

\begin{figure}[t]
    \centering
    \includegraphics[width=0.48\textwidth]{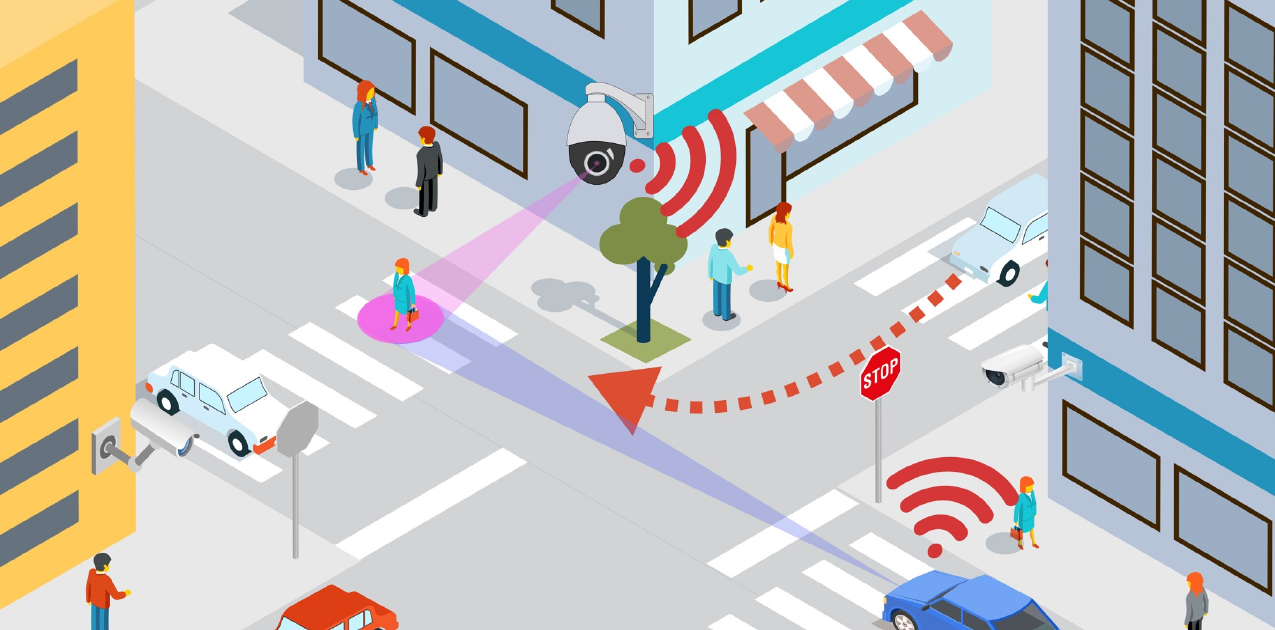}
    \caption{\revbis{Example of collective perception for road-safety enhancement.} A vehicle (or an RSU) detects a pedestrian and notifies its presence to an approaching car with limited visibility.}
    \vspace{-5pt}
    \label{fig:cps_scenario}
\end{figure}

In \emph{Day~2}, the paradigm shifts to \emph{``I share what I see''}, that is, to \emph{collective perception}~\cite{bib:WolfCPS2015,bib:WolfCPS2016}. In this phase, connected vehicles equipped with sensors such as cameras, lidars, and radars, broadcast information about their locally perceived environment. This information-sharing approach represents a powerful and transformative step toward enhanced road safety for several reasons:
\begin{itemize}
\item[(i)] Connected road users receiving perception messages can fuse the shared data with their own sensor measurements, thereby extending their perception range beyond the local sensor view;
\item[(ii)] Unconnected \acp{VRU} can be detected and their presence signaled to connected vehicles;
\item[(iii)] The effectiveness of collective perception is less dependent on the penetration rate of equipped vehicles compared to cooperative awareness, since vehicles or \acp{RSU} share what they perceive in the surroundings, including non-connected entities, enabling a broader situational awareness.
\end{itemize}
An example scenario where the Day~$2$ approach can significantly improve safety is illustrated in Fig.~\ref{fig:cps_scenario}.

Collective perception is currently one of the most prominent research and standardization themes in the road safety domain. In Europe, the format and generation rules of \acp{CPM} have been discussed for several years within \ac{ETSI}; however, the recently released standard is undergoing a substantial revision, as evidenced in the next section. Similar efforts are emerging globally: several pilot projects are underway in China; Korean technical reports refer to ``Extended Sensor Sharing Messages'' as the local equivalent of \acp{CPM}; and comparable initiatives 
are being developed in the United States.

Once Day~2 applications reach large-scale deployment, the next step will be \emph{Day~3}, characterized by the paradigm \emph{``I share what I plan to do and cooperate for safe and efficient maneuvers''}. This phase will be enabled through the exchange of messages called \acp{MCM} in Europe. As detailed in \ac{ETSI} TR~103~578~V2.1.1, \acp{MCM} will be used for agreement-seeking or prescribed operations: in the former case, vehicles negotiate their intended trajectories to reach a cooperative agreement, whereas in the latter they execute maneuvers prescribed by the infrastructure or another vehicle. A wide range of use cases can benefit from cooperative maneuvering, including lane merging and changing, intersection crossing, and platooning.

\section{Collective Perception}\label{sec:CP}

\subsection{The C-ITS Stack and the Collective Perception Service}
\label{subsection:ITS-stack}

As the TCP/IP suite is unfit for safety-of-life applications, the dedicated \ac{C-ITS} protocol stack was defined by \ac{ETSI} in EN 302 665. The stack includes three layers: from bottom to top, the access, networking \& transport, and facilities layer. The access layer is responsible for transmitting packets over the wireless medium, employing either the IEEE or 3GPP technologies. On top of it, the networking \& transport layer includes the \ac{GNW} protocol, which manages the addressing of nodes and the forwarding of packets in the case of multi-hop communications, and the \ac{BTP}, which is in charge of multiplexing the different flows received from the layer above. The last layer is the facilities layer, whose main components are the services. In \ac{C-ITS}, services are autonomous or application-controlled software entities that exchange data via structured messages. Unlike typical systems where applications handle data directly, \ac{C-ITS} services mainly generate and use data themselves and store it in dedicated databases, such as the local dynamic map. The main Day~1 services are the cooperative basic awareness service, which continuously generates \acp{CAM}, and the decentralized environmental notification service, which produces event-triggered \acp{DENM}. In the case of Day~2 collective perception, the facilities layer includes the \ac{CPS}, which continuously generates \acp{CPM}. The rules for message generation and object inclusion should balance a complex trade-off to ensure effectiveness while avoiding channel overload, as further discussed below.

\subsection{Collective Perception Messages}

\begin{figure*}[t]
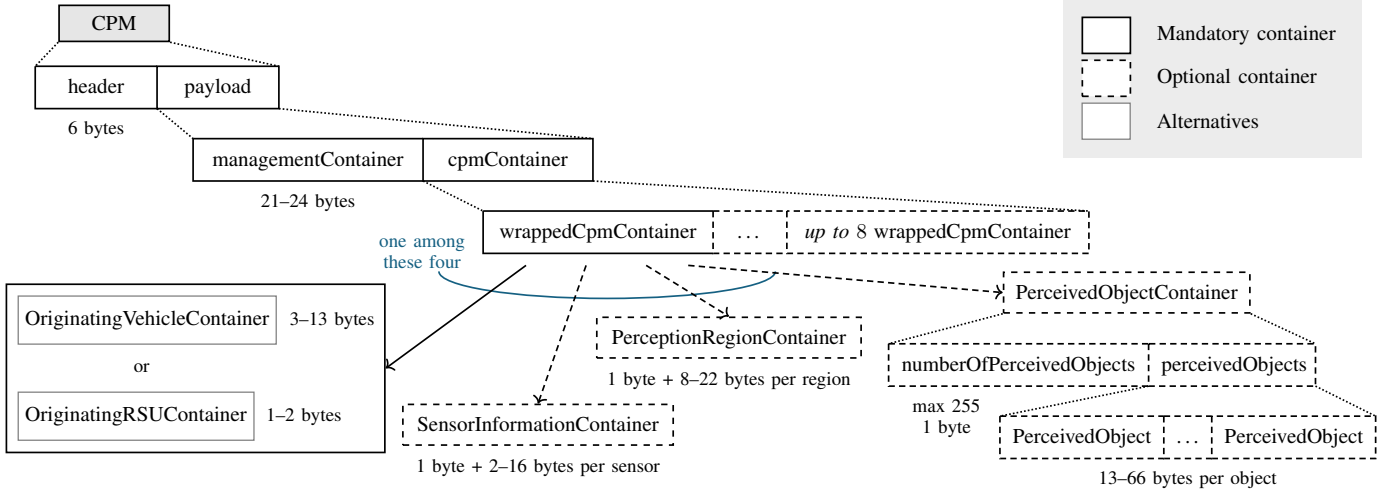

    \centering
    \includestandalone[width=\textwidth]{figures_R1/figure2_vittorio_v3}
    \caption{\revbis{Structure of the CPM defined by ETSI TS 103 324 V2.1.1 \cite{bib:ETSI_Cpm}, including the header, payload, and mandatory and optional containers.}}
    \label{fig:cpm_structure}
\end{figure*}

\ac{ETSI} mandates that the information shared by the \ac{CPS} is conveyed in the form of \acp{CPM}. The \ac{CPM} format includes a header and a payload made of several containers, as illustrated in Fig. \ref{fig:cpm_structure}. The management container provides the reference time and position of the transmitting station, along with segmentation data used when messages are fragmented. The \ac{CPM} container is a sequence of wrapped containers, each carrying different information. \rev{The first wrapped container describes the originating station}. For a vehicle, it reports the bounding-box dimensions, driving direction, speed, acceleration, and heading. For an \ac{RSU}, it provides the identifier of the intersection or road segment where it is located. Another wrapped container may include sensor information detailing the station's onboard sensing equipment: the number and types of sensors, identifiers, and detection areas. In a further wrapped container, an optional perception-region container specifies the deviations from the nominal sensing capabilities. One final wrapped container lists each perceived object with its identifier, type, and position, along with 
optional attributes such as the detecting sensors, confidence levels, age, and kinematic properties (e.g., speed, acceleration). This container is central to our analysis, as it largely determines the resulting \ac{CPM} size.

%


\subsection{The Issue of Redundancy Mitigation}

The transmission periodicity of \acp{CPM} ranges from $100~\text{ms}$ to $1~\text{s}$. According to \ac{ETSI} 103 324 \ac{CPM} specification, up to version 2.1.1, a message is transmitted when the prescribed object-inclusion conditions are met. \acp{VRU} are reported based on the time elapsed since the last inclusion in a message, whereas for vehicles and motorcyclists, the reporting also depends on their dynamic behavior, mirroring the logic used in \acp{CAM}. However, these choices may result in excessively large messages and a high channel load. For instance, vehicles in a congested area may perceive nearly the same large number of objects and redundantly report all of them in the messages. This issue motivated the study of redundancy mitigation techniques \cite{bib:Elche_Redundancy_Mitigation,bib:Festag_Redundancy_Mitigation}, which omit objects included too frequently, with limited motion, poor local confidence level, or too far from the disseminating station. Recently, the redundancy problem has prompted a substantial reconsideration of the message generation rules within \ac{ETSI}. 
\revbis{The early 2026 draft revision of ETSI TS~103~324 gives \ac{VoI} a prominent role in \ac{CPM} construction, indicating that objects with higher \ac{VoI} should be prioritized for inclusion in the message, while the number of advertised objects depends on the communication resources available for transmission.}
The \ac{VoI} of an object may depend on several factors, such as its proximity to the road user, the unpredictability of its motion, or the timeliness and reliability of its information. For instance, a pedestrian suddenly stepping onto the road, as in Fig. \ref{fig:cps_scenario}, has a much higher \ac{VoI} than a stationary parked car. By focusing on the most valuable data, vehicles and \acp{RSU} share fewer but more meaningful messages, reducing the load on the communication channel while improving collective situational awareness. However, different driving environments call for different interpretations of what the term ``valuable'' means. 
\rev{Computing the \ac{VoI} of an object is a challenging research question, since it is application-dependent, and its definition varies in different contexts \cite{bib:VoI_survey}. For vehicular networks, some 
proposals have recently emerged in literature~\cite{bib:Giordani_VoI, bib:schiegg1, bib:schiegg2}.}

\section{Results from Real Data}\label{sec:results}

In this section, we present the findings of our investigation into the capability of \ac{5G} \ac{NR}-\ac{V2X} to deliver collective perception traffic. To ensure a realistic evaluation, we employ object traces derived from real sensor measurements collected by moving vehicles, and conduct simulations using the open-source WiLabV2Xsim framework~\cite{bib:WiLabV2XSim}. The analysis focuses on a congested highway scenario in which a fleet of \ac{CPS}-capable vehicles share information about a large number of surrounding objects 
employing Mode $2$ \ac{SB-SPS}.

\subsection{Perceived Objects from Real Sensor Datasets}

\begin{figure}[t]
\centering



\subfloat[
\revbis{Example camera frame recorded in the Cirrus dataset \cite{bib:Cirrus_dataset_IEEE}.}
\label{fig:main-a}
]{%
    \includegraphics[
        width=\columnwidth,
        trim={0cm 1.2cm 0cm 0cm},
        clip
    ]{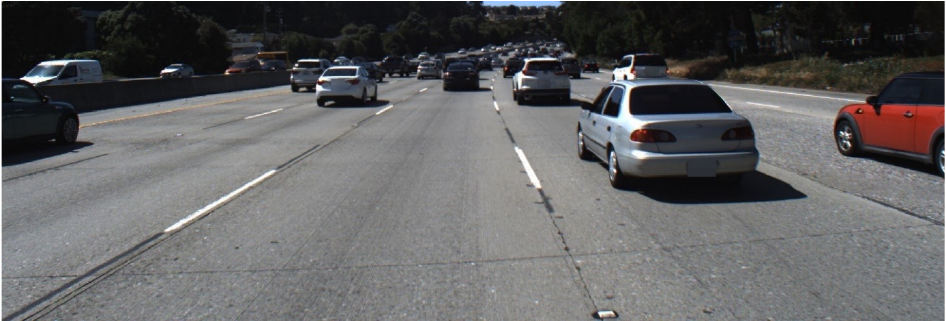}%
}

\vspace{0pt}

\captionsetup[subfloat]{captionskip=-0.75pt}
\subfloat[
Objects \revbis{included in a CPM as a function of time. $T_{\text{gen}}=200$ ms.}
\label{fig:main-b}
]{%
\makebox[\columnwidth][c]{%
\begin{tikzpicture}
\begin{axis}[
width=0.85\columnwidth,
height=0.3\columnwidth,
scale only axis,
xmin=0,
xmax=100,
xlabel={Time [s]},
xlabel style={yshift=2pt},
ymin=0,
ymax=65,
ylabel={Number of Objects},
axis background/.style={fill=white},
xmajorgrids,
ymajorgrids,
tick label style={font=\scriptsize},
label style={font=\scriptsize},
legend style={
  font=\scriptsize,
  row sep=-2pt,
  inner ysep=1pt,
  nodes={inner ysep=1pt},
  legend image post style={scale=0.5},
  legend cell align=left,
  align=left,
  at={(0.98,0.98)},
  anchor=north east,
  draw=white!15!black
}
]
\input{figures_R1/figure3b_200ms}
\end{axis}
\end{tikzpicture}%
}%
}

\vspace{-2pt}

\subfloat[
Packet size \revbis{and required subchannels} over time. $T_{\text{gen}}=200$ ms.
\label{fig:main-c}
]{%
\makebox[\columnwidth][c]{%
\begin{tikzpicture}
\begin{axis}[
width=0.85\columnwidth,
height=0.33\columnwidth,
scale only axis,
clip=false,
xmin=0,
xmax=100,
ymin=0,
ymax=1371,
xlabel={Time [s]},
xlabel style={yshift=2pt},
ylabel={Packet Size [B]},
xmajorgrids,
ymajorgrids,
axis background/.style={fill=white},
separate axis lines,
every outer y axis line/.append style={black},
every y tick label/.append style={font=\scriptsize\color{black}},
every x tick label/.append style={font=\scriptsize\color{black}},
every y tick/.append style={black},
every x tick/.append style={black},
label style={font=\scriptsize},
tick label style={font=\scriptsize},
yticklabel style={/pgf/number format/1000 sep={}},
legend style={
  font=\scriptsize,
  row sep=-2pt,
  inner ysep=1pt,
  nodes={inner ysep=1pt},
  legend image post style={scale=0.5},
  legend cell align=left,
  align=left,
  at={(0.93,0.98)},
  anchor=north east,
  draw=white!15!black
}
]
\input{figures_R1/figure3c_200ms}
\end{axis}
\end{tikzpicture}%
}%
}

\vspace{4pt}
\caption{\revbis{Starting from the object annotations in the dataset, the figure shows how different object-inclusion strategies affect the CPMs generated over time by the sensing vehicle. The number of included objects, packet sizes, and required  subchannels are derived from the real perception trace.}}
\label{fig:cirrus_morecongested}
\vspace{-12pt}
\end{figure}

A realistic characterization of collective perception traffic requires the use of actual sensor data. To this end, we employed the Cirrus dataset \cite{bib:Cirrus_dataset_IEEE}, an open-source collection of annotated scenes sampled at $1$~Hz by a vehicle driving through highways and urban streets located in Palo Alto, California. The recording vehicle was equipped with an RGB camera, two lidars,
two GPS units, and one IMU. An illustrative frame captured by the vehicle camera is shown in Fig.~\ref{fig:cirrus_morecongested}(a), referring to a congested highway scenario. \rev{Although Cirrus and similar datasets are typically used in computer vision tasks such as pattern and image recognition, here we leverage Cirrus content to address a different question: What is the number of perceived objects and their variability over time? This will serve as an input to characterize the perception traffic.}

In the dataset, a recorded scene consists of a sequence of annotated frames, and every annotation corresponds to an object represented by a bounding box. By tracking the evolution of the bounding boxes over time, the objects entering and leaving the vehicle’s field of view are identified. These correspond to the objects the vehicle detects~\cite{bib:cpm_dataset}, whose data may or may not be included in the \ac{CPM} depending on the examined message-generation strategy. Given the $1$~Hz sampling rate of the dataset, the frames 
provide the number of objects a vehicle would include when \acp{CPM} are generated every second. To investigate
shorter generation periods
(e.g., $200$ and $400$~ms), a linear interpolation of the number of \revbis{detected objects and their distance over time is applied.}

To determine the size of the packet transmitted over the radio channel, the relevant CPM containers shown in Fig.~\ref{fig:cpm_structure} are considered, then the ITS headers are added, as discussed in subsection \ref{subsection:ITS-stack}. Namely, the \ac{CPM} includes a $6$-byte header, a $21$-byte \emph{managementContainer}, a $5$-byte \emph{OriginatingVehicleContainer}, and a $10$-byte \emph{SensorInformationContainer} (updated every second). The \emph{PerceptionRegionContainer} is omitted. The \emph{PerceivedObjectContainer} contributes $13$ bytes per detected object, providing basic information. The \ac{BTP} and \ac{GNW} headers attached to the CPM
are 4 and 12 bytes long.
A $231$-byte full security certificate is further included in the packet every $1$ second; when the \ac{CPM} periodicity is lower than $1$~s, the $89$-byte digest is included in all remaining packets.


\subsection{Object Inclusion Strategies}
We examine three strategies for object inclusion: \begin{enumerate}
\item[(i)] \emph{Baseline}: all the detected objects are included in the \ac{CPM};
\item[(ii)]\emph{Random inclusion}: only a subset of the perceived objects is included in the message, with each object independently selected according to a probability $p_{\text{incl}}$, $p_{\text{incl}}<1$;
    \item[(iii)] \emph{\ac{VoI}-rank inclusion}: only the perceived objects with the highest \ac{VoI} value are included, subject to the constraint that the message does not exceed a predefined size, $S_{\text{max}}$.
\end{enumerate}

\medskip
The first strategy serves as the reference, while the second captures how reducing the \ac{CPM} size via the $p_{\text{incl}}$ value affects system performance. The third solution follows the new approach in the \ac{ETSI} \ac{CPS} standard, where the objects are ranked based on their \ac{VoI} and then those with the largest \ac{VoI} are selected until a maximum message size is reached; this maximum size is assumed fixed, but may be dynamically provided, for example, by a radio resource management entity such as the one that is under development in \ac{ETSI} \cite{10467184}. 
\revbis{Following \cite{bib:BLA-BLA}, given an object is at distance $d$ from the detecting vehicle, the \ac{VoI} of the object is computed according to a linear distance-based law with a negative slope as long as $d \leq \text{d}_{\text{max}}$, whereas it is $0$ if $d > \text{d}_{\text{max}}$, with $\text{d}_{\text{max}}= 50$~m. 
}   


\bigskip\bigskip
\bigskip
Communication performance is assessed in terms of: 
(i) \ac{PRR}, defined as the ratio between the number of receiving vehicles successfully decoding a packet at a given distance from the transmitting vehicle and the total number of vehicles located at that distance; (ii) \ac{PIR}, defined as the time between two consecutive successful receptions of packets  belonging to the same application flow \revbis{(as per ETSI TR 101 613)};
(iii) \ac{CBR}, defined as the fraction of subchannels \revtris{with a Received Signal Strength Indicator (RSSI) above -88~dBm (ETSI EN 303 798 - V2.1.1), within a 100 ms interval; this threshold is approximately 20~dB above the ideal thermal noise floor and indicates
the presence of strong interference;} 
we compute the median of the values obtained from all stations in all simulation intervals.
\revbis{To estimate perception effectiveness, we introduce the received VoI ratio (RVR). Following ETSI definition, the \textit{transmitted VoI} of a CPM is obtained as the sum of the VoI values of the objects included in the message. The corresponding \textit{received VoI} at distance $D$ is estimated as the transmitted VoI multiplied by the average PRR at that distance. 
The RVR is then obtained by averaging, over all CPMs, the received VoI normalized against the full-information case where all perceived objects are included every 100 ms and always correctly received.
Hence, the RVR represents the average fraction of the full-information reference that is delivered at a given distance. 
%
%
It captures, in a single metric, object selection, CPM generation frequency, and communication reliability.}

\subsection{Simulation Settings and Results}

In the simulated scenario, vehicles travel along a highway and are randomly distributed across six lanes, three in each direction. The average vehicular density is $150$ vehicles/km. \rev{The speed of each vehicle is drawn from a Gaussian distribution with mean $70$~km/h and standard deviation $7$ km/h; the mean corresponds to the estimated value in the congested setting recorded in the dataset.}
\rev{Moreover, all vehicles \revbis{are assumed to be capable of transmitting and receiving \acp{CPM}.}
The \ac{SL} channel is exclusively dedicated to CPS traffic, in alignment with envisioned real-world deployments.}

At physical layer, the transmission power is set to $23$~dBm, so that the power density depends on the number of subchannels the packet transmission requires. The antenna gain is $3$~dBi and the noise figure is $9$~dB. 
\rev{Following the ETSI indications, the \ac{SL} channel is $20$~MHz wide and the \ac{SCS} is $30$~kHz, with \revtris{at most} $5$ subchannels \revtris{available per slot to transmit a single packet (see Fig.~\ref{fig:cirrus_morecongested}(c), right axis)}. The \ac{MCS} index is set to $12$, which corresponds to $16$-QAM with coding rate $0.43$, and allows a maximum packet size of $1180$ bytes without fragmentation}. 
\rev{The propagation channel follows the ECC Report 68 rural model, recently adopted for highway scenarios in \ac{ETSI} TR 103 439.} 

We examine SB-SPS when the keep probability is set to $P=0.8$, and no retransmissions are considered.
 \rev{To handle the packet size variability, the allocation of subchannels matches the initial packet size. If the next packet requires fewer subchannels, padding is applied; if it requires more subchannels, a reallocation is performed.}

 \begin{figure*}[t]
    \centering
    \begin{tikzpicture}

\begin{groupplot}[
    group style={
        group size=3 by 1,
        horizontal sep=1.65cm,
        vertical sep=0pt
    },
    width=5.55cm,
    height=4.55cm,
    grid=both,
    major grid style={gray!30, solid},
    minor grid style={gray!10, dashed},
    xlabel={$D$ [m]},
    xmin=10,
    xmax=500,
    tick label style={font=\scriptsize},
    label style={font=\scriptsize},
    every axis plot/.append style={
        line width=2pt,
        mark=none
    }
]

\nextgroupplot[
    ylabel={PRR},
    ymin=0.85,
    ymax=1
]

\addplot [color=cpm100, dashed]
table [x index=0, y index=1, col sep=space]
{major_revision_2/tikzdata/bsln100.dat};

\addplot [color=cpm200]
table [x index=0, y index=1, col sep=space]
{major_revision_2/tikzdata/bsln200.dat};

\addplot [color=cpm400, dash dot]
table [x index=0, y index=1, col sep=space]
{major_revision_2/tikzdata/bsln400.dat};

\addplot [color=random]
table [x index=0, y index=1, col sep=space]
{major_revision_2/tikzdata/rnd200.dat};

\addplot [color=fixedtwo]
table [x index=0, y index=1, col sep=space]
{major_revision_2/tikzdata/voi200.dat};

\nextgroupplot[
    xlabel={$x$ [s]},
    ylabel={$\Pr\{\mathrm{PIR} \ge x\}$},
    xmin=0,
    xmax=2,
    ymin=0.0001,
    ymax=1,
    ymode=log,
    ytick={1e-5, 1e-4, 1e-3, 1e-2, 1e-1, 1},
    yticklabels={,$10^{-4}$,$10^{-3}$,$10^{-2}$,$10^{-1}$,$1$}
]

\addplot [color=cpm100, dashed]
table [x index=0, y index=1, col sep=space, each nth point=10]
{figures_R1/tikzdata/UpdateDelayComparison_semilog.dat};

\addplot [color=cpm200]
table [x index=0, y index=2, col sep=space, each nth point=10]
{figures_R1/tikzdata/UpdateDelayComparison_semilog.dat};

\addplot [color=cpm400, dash dot]
table [x index=0, y index=3, col sep=space, each nth point=10]
{figures_R1/tikzdata/UpdateDelayComparison_semilog.dat};

\addplot [color=random]
table [x index=0, y index=4, col sep=space, each nth point=10]
{figures_R1/tikzdata/UpdateDelayComparison_semilog.dat};

\addplot [color=fixedtwo]
table [x index=0, y index=5, col sep=space, each nth point=10]
{figures_R1/tikzdata/UpdateDelayComparison_semilog.dat};

\nextgroupplot[
    ylabel = {RVR},
    ymin=0,
    ymax=1
]

\addplot [color=cpm100, dashed]
table [x index=0, y index=1, col sep=space]
{major_revision_2/tikzdata/bsln100_avgrxvoi_linearDistance.dat};

\addplot [color=cpm200]
table [x index=0, y index=1, col sep=space]
{major_revision_2/tikzdata/bsln200_avgrxvoi_linearDistance.dat};

\addplot [color=cpm400, dash dot]
table [x index=0, y index=1, col sep=space]
{major_revision_2/tikzdata/bsln400_avgrxvoi_linearDistance.dat};

\addplot [color=random]
table [x index=0, y index=1, col sep=space]
{major_revision_2/tikzdata/rnd200_avgrxvoi_linearDistance.dat};

\addplot [color=fixedtwo]
table [x index=0, y index=1, col sep=space]
{major_revision_2/tikzdata/voi200_avgrxvoi_linearDistance.dat};

\end{groupplot}

\node[
    draw=white!15!black,
    fill=white,
    fill opacity=0.9,
    text opacity=1,
    inner sep=3pt
] at ($(current bounding box.north)+(0,0.55cm)$) {%
    \scriptsize
    \begin{tabular}{@{}c@{\hspace{14pt}}c@{\hspace{14pt}}c@{}}

    \tikz{\draw[cpm100, dashed, line width=2pt] (0,0) -- (0.42,0);}
    ~Baseline, $T_{\text{gen}}=100$ ms
    &
    \tikz{\draw[cpm200, line width=2pt] (0,0) -- (0.42,0);}
    ~Baseline, $T_{\text{gen}}=200$ ms
    &
    \tikz{\draw[cpm400, dash dot, line width=2pt] (0,0) -- (0.42,0);}
    ~Baseline, $T_{\text{gen}}=400$ ms
    \\[2pt]

    \multicolumn{3}{c}{
        \tikz{\draw[random, line width=2pt] (0,0) -- (0.42,0);}
        ~Random incl., $T_{\text{gen}}=200$ ms
        \hspace{18pt}
        \tikz{\draw[fixedtwo, line width=2pt] (0,0) -- (0.42,0);}
        ~VoI-rank, $T_{\text{gen}}=200$ ms
    }

    \end{tabular}
};

\node[
    font=\scriptsize,
    anchor=north
] at ($(group c1r1.south)+(0,-0.8cm)$)
{(a) Packet reception ratio};

\node[
    font=\scriptsize,
    anchor=north
] at ($(group c3r1.south)+(0,-0.8cm)$)
{\revbis{(c) Received VoI ratio}};

\node[
    font=\scriptsize,
    anchor=north
] at ($(group c2r1.south)+(0,-0.8cm)$)
{(b) PIR complementary CDF};

\end{tikzpicture}
    \vspace{-0.75cm}
    \caption{\revbis{Performance of the object-inclusion strategies: PRR,  CCDF of the packet inter-reception time, and received VoI ratio.}}
    \label{fig:prr_comparison}
\end{figure*}
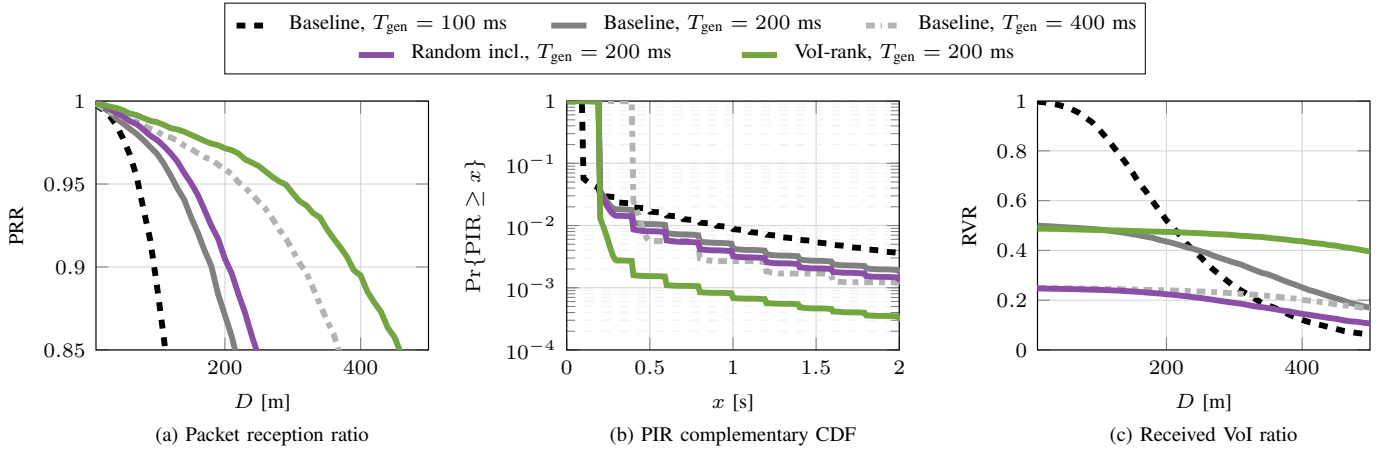

The temporal evolution of the number of objects and packet size is provided in Fig.~\ref{fig:cirrus_morecongested}. In particular, Fig.~\ref{fig:cirrus_morecongested}(b) showcases $N$, the number of objects included in each message as a function of time, for the three strategies: 
the \emph{baseline}, the \emph{random inclusion} when $p_{\text{incl}}=0.5$, and the \emph{VoI-rank inclusion}, when the latter employs a maximum packet size $S_{\text{max}}=400$ bytes\rev{ , corresponding to the largest size that fits in two subchannels}. The periodicity at which \acp{CPM} are generated is \rev{$T_{\text{gen}}=200$ ms}. Notice that for the \mbox{\emph{VoI-rank inclusion}} strategy, $N$ \rev{toggles between two values}: this is due to the transmission of the full security certificate every second, \rev{i.e., every five packets.} which intermittently limits the $N$ value. \rev{This phenomenon is not observed in the two other strategies, as they are not subject to the constant packet size constraint.} \rev{Fig.~\ref{fig:cirrus_morecongested}(c) reports the packet size evolution over time: for the VoI-rank inclusion scheme, it is equal to the maximum or differs from it for very few bytes. This is due to the congested highway conditions considered.}
On the right-hand side of this figure, the number of subchannels requested for packet transmission is shown, evidencing that it progressively reduces as stricter inclusion rules are introduced. Since the dataset contains objects from a single recording vehicle, each vehicle in the simulation is assigned a random starting point along the object trace. Once the end of the trace is reached, the sequence is read in reverse order, and this alternating forward–backward process continues until the end of the simulation.

\begin{figure}[t]
    \centering
    \begin{tikzpicture}

\definecolor{cpm100}{rgb}{0.00,0.00,0.00}
\definecolor{cpm200}{rgb}{0.50,0.50,0.50}
\definecolor{cpm400}{rgb}{0.70,0.70,0.70}
\definecolor{randincl}{rgb}{0.55,0.30,0.65}
\definecolor{voirank}{rgb}{0.45,0.65,0.25}

\begin{groupplot}[
    group style={
        group size=2 by 1,
        horizontal sep=0.1\columnwidth,
        vertical sep=0pt
    },
    width=0.45\columnwidth,
    height=0.30\columnwidth,
    scale only axis,
    xmin=0.4,
    xmax=5.6,
    ymin=0,
    xtick=\empty,
    xticklabels={},
    tick style={draw=none},
    yticklabel style={font=\scriptsize},
    tick label style={font=\scriptsize},
    label style={font=\scriptsize},
    title style={
        font=\scriptsize,
        at={(0.5,-0.18)},
        anchor=north,
        yshift=0pt,
        text height=1.6ex,
        text depth=.4ex
    },
    grid=major,
    major grid style={line width=.2pt, draw=gray!30},
    axis background/.style={fill=white},
    every axis plot/.append style={
        line width=0.3pt,
        mark=none
    }
]

\nextgroupplot[
    title={(a) Median CBR.},
    ymax=0.80
]

\addplot[
    ybar,
    bar width=11pt,
    mark=none,
    draw=cpm100,
    fill=cpm100,
    fill opacity=0.60
] coordinates {(1,0.768)};

\addplot[
    ybar,
    bar width=11pt,
    mark=none,
    draw=cpm200,
    fill=cpm200,
    fill opacity=0.60
] coordinates {(2,0.478)};

\addplot[
    ybar,
    bar width=11pt,
    mark=none,
    draw=cpm400,
    fill=cpm400,
    fill opacity=0.60
] coordinates {(3,0.238)};

\addplot[
    ybar,
    bar width=11pt,
    mark=none,
    draw=randincl,
    fill=randincl,
    fill opacity=0.60
] coordinates {(4,0.409)};

\addplot[
    ybar,
    bar width=11pt,
    mark=none,
    draw=voirank,
    fill=voirank,
    fill opacity=0.60
] coordinates {(5,0.323)};

\nextgroupplot[
    title={(b) Shared object perceptions per sec. \([\mathrm{s}^{-1}]\)},
    ymin=0,
    ymax=700,
    boxplot/draw direction=y,
    boxplot/box extend=0.55,
    boxplot/every whisker/.style={draw=black}
]

\addplot[
    mark=none,
    draw=cpm100,
    fill=cpm100,
    fill opacity=0.55,
    boxplot prepared={
        lower whisker=171,
        lower quartile=245.5,
        median=288,
        upper quartile=397.5,
        upper whisker=611
    }
] coordinates {};

\addplot[
    mark=none,
    draw=cpm200,
    fill=cpm200,
    fill opacity=0.55,
    boxplot prepared={
        lower whisker=86,
        lower quartile=122.5,
        median=143,
        upper quartile=199.5,
        upper whisker=306
    }
] coordinates {};

\addplot[
    mark=none,
    draw=cpm400,
    fill=cpm400,
    fill opacity=0.55,
    boxplot prepared={
        lower whisker=52,
        lower quartile=72.5,
        median=87,
        upper quartile=121.25,
        upper whisker=183
    }
] coordinates {};

\addplot[
    mark=none,
    draw=randincl,
    fill=randincl,
    fill opacity=0.55,
    boxplot prepared={
        lower whisker=37,
        lower quartile=58.5,
        median=71,
        upper quartile=101,
        upper whisker=155
    }
] coordinates {};

\addplot[
    mark=none,
    draw=voirank,
    fill=voirank,
    fill opacity=0.55,
    boxplot prepared={
        lower whisker=75,
        lower quartile=75,
        median=75,
        upper quartile=75,
        upper whisker=75
    }
] coordinates {};

\end{groupplot}

\node[
    draw=black,
    fill=white,
    inner sep=3pt,
    anchor=south
] at ($(current bounding box.north)+(0,1.0em)$) {%
    \scriptsize
    \begin{tabular}{@{}c@{\hspace{10pt}}c@{\hspace{10pt}}c@{}}

    \tikz{\draw[cpm100, fill=cpm100, fill opacity=0.60]
    (0,0) rectangle (0.22,0.12);}
    ~Baseline, $T_{\text{gen}}=100$~ms
    &
    \tikz{\draw[cpm200, fill=cpm200, fill opacity=0.60]
    (0,0) rectangle (0.22,0.12);}
    ~Baseline, $T_{\text{gen}}=200$~ms
    &
    \tikz{\draw[cpm400, fill=cpm400, fill opacity=0.60]
    (0,0) rectangle (0.22,0.12);}
    ~Baseline, $T_{\text{gen}}=400$~ms
    \\[1pt]

    \multicolumn{3}{c}{%
        \tikz{\draw[randincl, fill=randincl, fill opacity=0.60]
        (0,0) rectangle (0.22,0.12);}
        ~Random incl., $T_{\text{gen}}=200$~ms
        \hspace{16pt}
        \tikz{\draw[voirank, fill=voirank, fill opacity=0.60]
        (0,0) rectangle (0.22,0.12);}
        ~VoI-rank, $T_{\text{gen}}=200$~ms
    }

    \end{tabular}
};

\end{tikzpicture}
    \caption{\revbis{Channel busy ratio and shared objects per second.} 
    }
    \label{fig:othermetrics_comparison}
    \vspace{-7pt}
\end{figure}
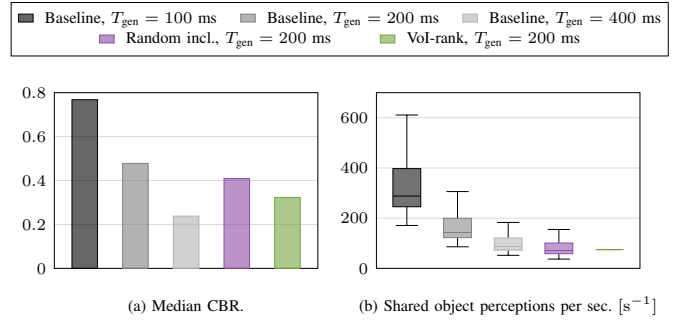

\revbis{The impact of communication is addressed in Fig.~\ref{fig:prr_comparison}. 
Fig.~\ref{fig:prr_comparison}(a) reports the \ac{PRR} as a function of the transmitter--receiver distance for the considered object-inclusion strategies. The baseline with $T_{\text{gen}}=100$~ms disseminates all perceived objects most frequently, but exhibits the fastest PRR degradation with distance. Increasing $T_{\text{gen}}$ to $200$ or $400$~ms improves the PRR range at the price of less frequent perception updates, while the VoI-rank strategy with $T_{\text{gen}}=200$~ms further benefits from a limited and stable packet size.}
\revbis{Fig.~\ref{fig:prr_comparison}(b) complements this view through the \ac{PIR} \ac{CCDF} for distances below $300$~m. The results show the trade-off between frequent updates and reception gaps, with the VoI-rank strategy limiting the probability of large PIR values.}
\revbis{Fig.~\ref{fig:prr_comparison}(c) reports the RVR, which combines reliability with the value of the delivered perception content. The $100$~ms baseline achieves the highest RVR close to the transmitter, where the PRR is high and all perceived objects are included at the reference generation period ($100$~ms), but its value decreases rapidly with distance. The $200$~ms and $400$~ms baselines are penalized by their lower generation frequency, while random inclusion is additionally affected by the reduced transmitted VoI. The VoI-rank strategy provides a more stable RVR over distance, since it limits the message size while prioritizing objects with larger VoI.}

\revbis{Finally, Fig.~\ref{fig:othermetrics_comparison} provides additional information in terms of channel occupancy and shared objects. The median \ac{CBR} in Fig.~\ref{fig:othermetrics_comparison}(a) confirms that less frequent or smaller CPMs reduce channel congestion. Conversely, Fig.~\ref{fig:othermetrics_comparison}(b) shows that reducing the channel load generally also reduces the number of objects shared per second, confirming that the object selection should be guided by a relevance criterion.}

\revbis{Overall, the results show that the most suitable strategy depends on the balance among selected information value, message generation frequency, communication reliability, and channel occupancy.}

\section{Conclusion and Future Directions}\label{sec:conclusion}
\revbis{This work reviewed the status of \ac{CPS} standardization, investigated the impact of \ac{CPM} traffic on communication performance, and evaluated the resulting perception effectiveness, considering \ac{5G}~\ac{NR}-\ac{V2X} \ac{SL} as the underlying communication technology.}
Real-world object traces were integrated into a network-level simulator to first assess communication reliability, \rev{timeliness}, and resource occupancy, \rev{when different strategies for including object information in the CPMs are employed.}
\revbis{The conventional communication metrics were then paired by a simple estimate of the amount of valuable information that CPMs convey to receiving vehicles.}
\rev{
The results
indicate
that variations in message size significantly affect the performance of 5G NR-V2X \ac{SL} and that the VoI-rank inclusion strategy yields substantial \revbis{communication} gains by shaping perception traffic in line with current ETSI standardization. 
\revbis{The latter scheme also exhibits a remarkable delivery of valuable information, with limited loss increasing the distance.}
In \revbis{the proposed} approach, the access layer notifies resource availability in terms of packet size, thereby constraining the message dimension at the facility layer, while the CPS independently ranks objects based on their relevance and fills messages accordingly, thus simplifying system design.
}

\revbis{Several avenues remain for future research.}
First, the optimal message size should be determined for different configurations of the physical and access layer, and under different CPM traffic conditions. As a second point, the definition of the value of information assigned to the perceived objects \revbis{requires additional effort to accurately capture a broader range of aspects, including accuracy, timeliness, and relevance to the receivers.} 
Moreover,
future research necessitates novel datasets to quantify performance in alternative settings, such as urban areas.
Finally, large-scale field experiments are still required before practical deployment can be pursued.

\section*{Acknowledgments}
This work was supported by the Italian National projects MoVeOver and ALERT of the ``RESTART'' program (PE00000001) and by the DIEF FAR 2025 research grant. 
The authors sincerely thank the reviewers for their insightful comments, which greatly enhanced the quality of this work.

\bibliographystyle{IEEEtran}
\bibliography{bib_folder/IEEEabrv,bib_folder/bibliography.bib}

\begin{thebibliography}{10}
\providecommand{\url}[1]{#1}
\csname url@samestyle\endcsname
\providecommand{\newblock}{\relax}
\providecommand{\bibinfo}[2]{#2}
\providecommand{\BIBentrySTDinterwordspacing}{\spaceskip=0pt\relax}
\providecommand{\BIBentryALTinterwordstretchfactor}{4}
\providecommand{\BIBentryALTinterwordspacing}{\spaceskip=\fontdimen2\font plus
\BIBentryALTinterwordstretchfactor\fontdimen3\font minus
  \fontdimen4\font\relax}
\providecommand{\BIBforeignlanguage}[2]{{%
\expandafter\ifx\csname l@#1\endcsname\relax
\typeout{** WARNING: IEEEtran.bst: No hyphenation pattern has been}%
\typeout{** loaded for the language `#1'. Using the pattern for}%
\typeout{** the default language instead.}%
\else
\language=\csname l@#1\endcsname
\fi
#2}}
\providecommand{\BIBdecl}{\relax}
\BIBdecl

\bibitem{bib:WolfCPS2015}
H.-J. Gunther, O.~Trauer, and L.~Wolf, ``The potential of collective perception
  in vehicular ad-hoc networks,'' in \emph{ITST}, 2015.

\bibitem{bib:WolfCPS2016}
H.-J. Günther \emph{et~al.}, ``Realizing collective perception in a vehicle,''
  in \emph{IEEE VNC}, 2016.

\bibitem{bib:Elche_Redundancy_Mitigation}
G.~Thandavarayan, M.~Sepulcre, and J.~Gozalvez, ``Redundancy mitigation in
  cooperative perception for connected and automated vehicles,'' in \emph{IEEE
  VTC2020-Spring}, 2020.

\bibitem{bib:Festag_Redundancy_Mitigation}
Q.~Delooz \emph{et~al.}, ``Analysis and evaluation of information redundancy
  mitigation for {V2X} collective perception,'' \emph{IEEE Access}, vol.~10,
  pp. 47\,076--47\,093, 2022.

\bibitem{bib:evolutionV2Xstandards}
G.~Naik, B.~Choudhury, and J.-M. Park, ``{IEEE 802.11bd} \& {5G NR V2X}:
  Evolution of radio access technologies for {V2X} communications,'' \emph{IEEE
  Access}, vol.~7, pp. 70\,169--70\,184, 2019.

\bibitem{bib:ETSI_Cpm}
{ETSI}, ``{Intelligent Transport System (ITS); Vehicular Communications; Basic
  Set of Applications; Collective Perception Service; Release 2},'' Tech. Spec.
  103 324 V2.1.1, Jun 2023.

\bibitem{bib:VoI_survey}
F.~Alawad and F.~A. Kraemer, ``Value of information in wireless sensor network
  applications and the {IoT}: A review,'' \emph{IEEE Sensors Journal}, vol.~22,
  no.~10, pp. 9228--9245, 2022.

\bibitem{bib:Giordani_VoI}
M.~Giordani \emph{et~al.}, ``A framework to assess value of information in
  future vehicular networks,'' in \emph{Proceedings of the 1st ACM MobiHoc
  Workshop on Technologies, MOdels, and Protocols for Cooperative Connected
  Cars}, 2019.

\bibitem{bib:schiegg1}
T.~Lyu \emph{et~al.}, ``Accuracy and relevance: A value of information based
  prioritisation of perceived objects for the etsi collective perception
  service,'' in \emph{IEEE VNC}, 2025.

\bibitem{bib:schiegg2}
V.~A. Wolff \emph{et~al.}, ``Uncertainty and prioritization: Empirical
  evaluation of a {VoI}-based {CPM} generation pipeline using real-world
  data,'' in \emph{IEEE VNC}, 2025.

\bibitem{bib:WiLabV2XSim}
V.~Todisco \emph{et~al.}, ``Performance analysis of sidelink {5G-V2X} mode 2
  through an open-source simulator,'' \emph{IEEE Access}, vol.~9, pp.
  145\,648--145\,661, 2021.

\bibitem{bib:Cirrus_dataset_IEEE}
Z.~Wang \emph{et~al.}, ``Cirrus: A long-range bi-pattern {LiDAR} dataset,'' in
  \emph{IEEE ICRA}, 2021.

\bibitem{bib:cpm_dataset}
M.~Andreani, L.~Lusvarghi, and M.~L. Merani, ``A statistical characterization
  of the actual cooperative perception messages and a generative model to
  reproduce them,'' in \emph{IEEE FNWF}, 2023.

\bibitem{10467184}
A.~Bazzi \emph{et~al.}, ``Multi-channel operation for the release 2 of {ETSI}
  cooperative intelligent transport systems,'' \emph{IEEE Communications
  Standards Magazine}, vol.~8, no.~1, pp. 28--35, 2024.

\bibitem{bib:BLA-BLA}
M.~Sepulcre, J.~Tortosa-Garcia, and J.~Gozalvez, ``Demand- and priority-aware
  adaptive congestion control for heterogeneous {V2X} service requirements,''
  in \emph{IEEE VTC2026-Spring}, 2026.

\end{thebibliography}

\vspace{-2.5em}

\begin{IEEEbiographynophoto}{Vittorio Todisco} [M] (vittorio.todisco@unibo.it) is a research fellow at the University of Bologna, where he obtained his Ph.D. His work focuses on connected vehicles and standardization activities in the ITS sector.
\end{IEEEbiographynophoto}
\vspace{-3em}

\begin{IEEEbiographynophoto}{Mattia Andreani}
[GSM] (mattia.andreani@unimore.it) is a Ph.D. student in ICT at the University of Modena and Reggio Emilia. His research interests include vehicular communications and vehicular collective perception. 
\end{IEEEbiographynophoto}
\vspace{-3em}

\begin{IEEEbiographynophoto}{Maria Luisa Merani}
[SM] (marialuisa.merani@unimore.it) is an Associate Professor at the University of Modena and Reggio Emilia. Her research interests focus on vehicular networking and road safety. She served as an Editor of the IEEE Transactions on Wireless Communications and as General Chair of the 2025 IEEE WCNC conference. 
\end{IEEEbiographynophoto}
\vspace{-3em}

\begin{IEEEbiographynophoto}{Alessandro Bazzi}
[SM] (alessandro.bazzi@unibo.it) is an Associate Professor at the University of Bologna. His research interests are mainly on medium access control and radio resource management of networks of connected and autonomous vehicles. He is/had been part of the ETSI STFs on C-ITS. 
\end{IEEEbiographynophoto}

\end{document}